\documentclass[12pt,a4paper]{article}

\usepackage{amsmath,amssymb}
\usepackage{bm}
\usepackage{graphicx}
\usepackage{ascmac}
\usepackage{wrapfig}
\usepackage{braket}
\usepackage{color}
\usepackage{mathrsfs}
\usepackage[english]{babel}
\usepackage{inconsolata}
\usepackage{cases}
\usepackage{tikz}
\usepackage{cite}
\usepackage{setspace}
\usetikzlibrary{intersections, calc}
\usepackage[top=3cm, bottom=2cm, left=2.3cm, right=2.3cm, includefoot]{geometry}

\newcommand{\beq}{\begin{equation}}
\newcommand{\eeq}{\end{equation}}
\newcommand{\lr}[1]{\left(#1\right)}

\newcommand{\Schro}{Schr$\rm{\ddot{o}}$dinger\ }

\makeatletter
\@addtoreset{equation}{section}

\makeatother

\begin{document}

\begin{titlepage}

\pagenumbering{gobble}

\begin{flushright}
\rm TIT/HEP-679 \\ February, 2020
\end{flushright}

\renewcommand{\thefootnote}{*}

\vspace{0.2in}
\begin{center}
\Large{\bf Exact WKB analysis and TBA equations \\ for the Mathieu equation}
\end{center}
\vspace{0.2in}
\begin{center}
\large Keita Imaizumi\footnote{E-mail: k.imaizumi@th.phys.titech.ac.jp}
\end{center}

\begin{center}{\it
Department of Physics,
\par
Tokyo Institute of Technology
\par
Tokyo, 152-8551, Japan
}
\end{center}
\vspace{0.2in}
\begin{abstract}
\begin{spacing}{1.0}
{\footnotesize We derive the Thermodynamic Bethe Ansatz (TBA) equations for the exact WKB periods of the Mathieu equation in the weak coupling region. We will use the TBA equations to calculate the quantum corrections to the WKB periods, which are regarded as the quantum periods of $\mathcal{N} = 2$ $SU(2)$ super Yang-Mills theory at strong coupling. We calculate the effective central charge of the TBA equations, which is found to be proportional to the coefficient of the one-loop beta function of the 4d theory. We also study the spectral problem for the Mathieu equation based on the TBA equations numerically. }
\end{spacing}
\end{abstract}

\end{titlepage}

\newpage

\pagenumbering{arabic}
\setcounter{page}{1}

\renewcommand{\thefootnote}{\arabic{footnote}}
\setcounter{footnote}{0}

\section{Introduction}
\label{sec:intro}
The exact WKB analysis for the one-dimensional \Schro equation provide a rigorous approach to the study of the spectral problem \cite{Vor,devet,DP,Esemi}. In this analysis, the \Schro equation is formulated in a complex plane. The WKB periods are asymptotic series in the Planck constant $\hbar$, which is also regarded as a complex parameter. The WKB periods have the discontinuities in the complex $\hbar$-plane. In \cite{Vor}, the author has pointed out that their discontinuity structure and their classical limit determine the WKB periods completely (analytic bootstrap program \cite{Vor}). Recently, the solution to this problem has been explicitly derived for arbitrary polynomial potentials \cite{Ito}, and for potentials with a regular singularity \cite{IS}. The solutions in \cite{Ito,IS} take the form of the Thermodynamic Bethe Ansatz equations (for short, the TBA equations), which are the integral equations that appear in the study of the integrable field theories represented by 2-dimensional CFT.
\par
An interesting generalization is the case of the periodic potential. In particular, the Mathieu equation has been studied in detail using the exact WKB analysis \cite{Grassi,Kashani}. In \cite{Grassi}, the Borel transformation to the WKB periods for the modified version of the Mathieu equation are shown to agree numerically with the result of the TBA equations \cite{GMN,Opers}. 
\par
The spectral problem for the Mathieu equation has been also investigated as an example of the theories which have an infinite number of vacua (e.g. \cite{BD,DU}). The energy spectrum of the Mathieu equation has the band structure, in which the non-perturbative effects contribute to the bandwidth. Moreover, in the weak coupling regime, the one-instanton contribution for the energy spectrum can be expressed in terms of the perturbative WKB periods \cite{Res1,Res2}. This shows a specific example of the "resurgence structure" between perturbative sectors and non-perturbative sectors.
\par
The WKB periods for the Mathieu equation can also be regarded as the quantum periods, which are the Seiberg-Witten periods \cite{SW} in the Nekrasov-Shatashivili limit of the $\Omega$-background, for 4-dimensional $\mathcal{N} = 2$ $SU(2)$ super Yang-Mills theory \cite{MM,NS}. This relation is used to the calculation of the instanton correction to the Nekrasov partition function \cite{Kre}. And in \cite{Kashani}, the instanton corrections to the prepotential of $\mathcal{N}=2$ SU(2) SYM has been computed by using the exact WKB analysis. The quantum SW curve for $\mathcal{N}=2$ gauge theories have been studied in \cite{IOK,Okubo1,Okubo2}.
\par
The purpose of the present paper is to derive the TBA equations for the exact WKB periods of the Mathieu equation in the weak coupling region. We will use the TBA equations to calculate the quantum corrections to the WKB periods, which are regarded the quantum periods of $\mathcal{N} = 2$ $SU(2)$ super Yang-Mills theory at strong coupling. We calculate the effective central charge of the TBA equations, which is found to be proportional to the coefficient of the one-loop beta function of the 4d theory. We also study the spectral problem for the Mathieu equation based on the TBA equations numerically. 
\par
This paper is organized as follows. In section \ref{sec:EWAM}, we apply the exact WKB analysis to the Mathieu equation and compute the discontinuities for the WKB periods. In section \ref{sec:MTBA}, using the discontinuities, we derive the TBA equations for the Mathieu equation. This TBA equation clarify non-trivial relations between the 4-dimensional QFT and the underlying 2-dimensional CFT. We also present some numerical results of the TBA equations. Finally, in section \ref{sec:band}, we apply the TBA equations to study the energy spectrum of the Mathieu equation numerically.

\section{Exact WKB analysis for the Mathieu equation}
\label{sec:EWAM}
We consider the Mathieu equation\footnote{Our analysis is converted to the modified Mathieu equation \cite{Grassi} by the change of variables $q \rightarrow ix$, $\hbar \rightarrow i\hbar$.} on the Riemann sphere $\mathbb{C}^{\ast} = \mathbb{C} \cup \{\infty\}$,
\begin{equation}
\label{eq:Meq}
 \left(-\hbar^2\frac{d^2}{dq^2} + \Lambda^2\cos q\right)\phi(q) = u\phi(q),
\end{equation}
where $\Lambda \in \mathbb{R}$, $u, \hbar \in \mathbb{C}$ and $q$ is a complex coordinate on $\mathbb{C}^{\ast}$.
\par
In the standard WKB method, we assume that the wave function $\phi(q)$ has the following asymptotic expansion in $\hbar$,
\beq
\label{eq:WKBS}
 \phi(q) = \frac{1}{\sqrt{P(q)}}\exp(\frac{i}{\hbar}\int^{q}P(q')dq'),
\eeq
\begin{equation}
  P(q) = \sum_{n=0}^{\infty}p_{n}(q)\hbar^{2n}.
\end{equation}
We can obtain $p_{n}$ recursively by substituting (\ref{eq:WKBS}) into (\ref{eq:Meq}). In particular, $p_0(q) = \sqrt{u - \Lambda^2\cos(q)}$ is the classical momentum. 
\par
The one-form $P(q)dq$ can be regarded as a meromorphic differential on the curve
\begin{equation}
 y^2 = u - \Lambda^2\cos(q).
\end{equation}
This curve defines a Riemann surface $\hat{\Sigma}$, which we will call WKB curve. The WKB curve $\hat{\Sigma}$ is a branched covering of $\mathbb{C}^{\ast}$. Along the one-cycles $\gamma \in H_1(\hat{\Sigma})$, we can define the period integrals of $P(q)dq$, which we will call "standard" WKB periods,
\begin{equation}
a_{\gamma} := \oint_{\gamma} P(q)dq, \ \ \ \ \ \ \gamma \in H_1(\hat{\Sigma}).
\end{equation}
These periods are expanded as the power series in $\hbar$,
\begin{equation}
  a_{\gamma} = \sum_{n=0}^{\infty} a_{\gamma}^{(n)}\hbar^{2n},\ \ \ \ \ \ a_{\gamma}^{(n)} = \oint_{\gamma}p_n(q)dq. 
\end{equation}
In particular, we will call $a_{\gamma}^{(0)}$ classical periods, which are the integrals of the classical momentum along $\gamma$. 
\par
The standard WKB periods are an asymptotic series in $\hbar$. Then we define the Borel transformation,
\begin{equation}
 \hat{a}_{\gamma}(\xi) = \sum_{n=0}^{\infty}\frac{1}{(2n)!}a_{\gamma}^{(n)}\xi^{2n}, \ \ \ (\xi \in \mathbb{C})
\end{equation}
and the Borel resummation,
\begin{equation}
\label{eq:Bsum}
 s_{\varphi}\left(a_{\gamma}\right)\left(\hbar\right) = \frac{1}{\hbar}\int_{0}^{e^{i\varphi}\infty} e^{-\xi/\hbar}\hat{a}_{\gamma}(\xi)d\xi,
\end{equation}
where $\varphi$ is the phase of $\hbar$ ($\hbar = |\hbar|e^{i\varphi}$). We can obtain an analytic function $s_{\varphi}\left(a_{\gamma}\right)\left(\hbar\right)$, from which we recover the standard WKB periods $a_{\gamma}$ as the asymptotic expansion in $\hbar$. In this paper, we refer to the analytic continuation of $s_{\varphi}\left(a_{\gamma}\right)\left(\hbar\right)$ as "exact" WKB periods. The Borel summability of the standard WKB periods for a class of meromorphic potentials has been discussed in \cite{cr1,cr2}.
\par
The analytic continuations of the Borel transformations can have some poles and branch cuts in the $\xi$-plane. To avoid the singularities in the integration contour of (\ref{eq:Bsum}), we define the lateral Borel resummation,
\begin{equation}
 s_{\varphi\pm}\left(a_{\gamma}\right)\left(e^{i\varphi}|\hbar|\right) = \lim_{\delta \rightarrow +0} s_{\varphi \pm \delta}\left(a_{\gamma}\right)\left(e^{i(\varphi \pm \delta)}|\hbar|\right).
\end{equation}
 $s_{\varphi}\left(a_{\gamma}\right)$ has discontinuity along the direction $\varphi$, which is can be expressed as
\begin{equation}
 \mathrm{disc}_{\varphi} a_{\gamma} := s_{\varphi+}\left(a_{\gamma}\right) - s_{\varphi-}\left(a_{\gamma}\right). 
\end{equation}
\par
Let us now calculate the discontinuities of the exact WKB periods for the Mathieu equation (\ref{eq:Meq}). First, we define the canonical one-cycles on $\hat{\Sigma}$. On the complex $q$-plane, we define the branch cuts between the turning points. We use the coordinate with the identification $q \sim q + 4\pi$ while the potential $\cos q$ has the periodicity $2\pi$. This choice is useful to formulate the TBA equations as we will see. In this coordinate, the complex plane is compactified by adding the points at ${\pm i\infty}$ and these points are not the branch points. This is in contrast with the case of \cite{Kashani}, where the coordinate $q$ identified as $q \sim q + 2\pi$. In this coordinate, the points ${\pm i\infty}$ become the branch points and one needs to consider the cuts between the turning point and infinity.
\par
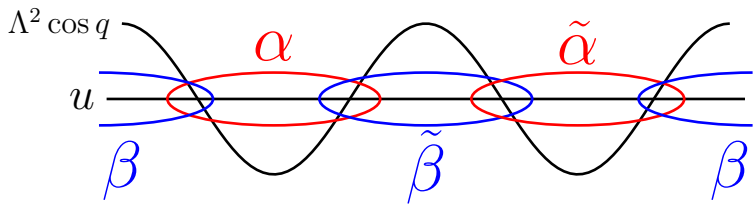
\begin{figure}[htbp]
  \begin{center}
  \hspace*{-0.2in}
   \begin{tikzpicture}
    \draw[domain=4.2:-4.2,line width=1pt]plot(\x,0) node[left]{\Large$u$};
    \draw[samples=100,domain=4:-4,line width=1pt] plot(\x,{cos(pi*\x/2 r)}) node[left]{$\Lambda^2\cos q$};
    \draw[red, line width=1pt] (-2,0) circle [x radius=1.4cm, y radius=3.5mm] node at (-2,0.7) {\huge$\alpha$};
    \draw[blue, line width=1pt] (0,0) circle [x radius=1.4cm, y radius=3.5mm] node at (0,-0.9) {\huge$\tilde{\beta}$};
    \draw[red, line width=1pt] (2,0) circle [x radius=1.4cm, y radius=3.5mm] node at (2,0.75) {\huge$\tilde{\alpha}$};
    \draw[blue, line width=1pt] (-2.8,0) arc (0:90: 1.5cm and 0.35cm);
    \draw[blue, line width=1pt] (-4.3,-0.35) arc (270:360: 1.5cm and 0.35cm)node at (-4,-0.95) {\huge$\beta$};
    \draw[blue, line width=1pt] (2.8,0) arc (180:90: 1.5cm and 0.35cm);
    \draw[blue, line width=1pt] (2.8,0) arc (180:270: 1.5cm and 0.35cm)node at (4,-0.95) {\huge$\beta$};
   \end{tikzpicture}
   
   \caption{The branch points and cycles with the identification $q \sim q + 4\pi$.}
   \label{fig:cocycle}
  \end{center}
\end{figure}

   
In the following analysis, we will restrict to the parameter region as $-\Lambda^2 < u < \Lambda^2$, which correspond to the weak coupling region of the Mathieu equation. This parameter region is called the minimal chamber \cite{MC}. In this regime, we can take the WKB curve $\hat{\Sigma}$ so that the branch cuts lie on the classically allowed intervals. These branch cuts define four independent cycles, the cycles $\alpha$ and $\tilde{\alpha}$ which encircle the classically allowed intervals, and $\beta$, $\tilde{\beta}$ which encircle the classically forbidden intervals (Fig.\ref{fig:cocycle}). 

We choose the orientations of the cycles in such a way that the classical periods,
\begin{equation}
\begin{split}
 & m_1 := a^{(0)}_{\alpha} = \oint_{\alpha} p_0(q)dq, \ \ \ m_2 := ia_{\beta}^{(0)} = i\oint_{\beta}p_0(q)dq, \\
 & m_3 := a^{(0)}_{\tilde{\alpha}} = \oint_{\tilde{\alpha}} p_0(q)dq, \ \ \ m_4 := ia_{\tilde{\beta}}^{(0)} = i\oint_{\tilde{\beta}}p_0(q)dq,
\end{split}
\end{equation}
are real and positive. In fact, from the periodicity of $\cos q$, the periods satisfy 
\beq
\label{eq:id}
 a_{\alpha} = a_{\tilde{\alpha}}, \ \ \ a_{\beta} = a_{\tilde{\beta}}.
\eeq
Thus we take $a_{\alpha}$, $a_{\beta}$ as the independent periods for the Mathieu equation.
\par
The discontinuities of the exact periods can be captured by the Delabaere-Pham formula (Theorem 2.5.1 of \cite{DP}, and Theorem 3.4 of \cite{cr1}). The Delabaere-Pham formula says that the exact periods for the classically allowed intervals have the discontinuities in the directions for $\varphi = 0$ in the complex $\hbar$-plane, and its discontinuities are determined by the exact periods for the classically forbidden intervals which cycles intersect with the each classically allowed intervals,
\begin{equation}
\label{eq:adisc}
 \frac{i}{\hbar}\mathrm{disc}_{0}\ a_{\alpha} = -\log\left[1 + \exp\left(-\frac{i}{|\hbar|}s_0(a_{\beta})\right)\right] -\log\left[1 + \exp\left(-\frac{i}{|\hbar|}s_0(a_{\tilde{\beta}})\right)\right].
\end{equation}
The r.h.s. of (\ref{eq:adisc}) is rearranged by the identification (\ref{eq:id}) as
\begin{equation}
\label{eq:adisc1}
 \frac{i}{\hbar}\mathrm{disc}_{0}\ a_{\alpha} = -2\log\left[1 + \exp\left(-\frac{i}{|\hbar|}s_0(a_{\beta})\right)\right].
\end{equation}
The similar formula hold for the directions for $\varphi = \pi$,
\begin{equation}
\label{eq:adisc2}
 \frac{i}{\hbar}\mathrm{disc}_{\pi}\ a_{\alpha} = +2\log\left[1 + \exp\left(-\frac{i}{|\hbar|}s_0(a_{\beta})\right)\right].
\end{equation}
In contrast, the exact periods for the classically forbidden intervals have the discontinuities in the directions for $\varphi = \pm \frac{\pi}{2}$, and its discontinuities are determined by the exact periods for the classically allowed intervals which cycles intersect with the each classically forbidden intervals,
\begin{equation}
\label{eq:bdisc}
 \frac{i}{\hbar}\mathrm{disc}_{\pm\frac{\pi}{2}}\ a_{\beta} = \pm 2\log\left[1 + \exp\left(-\frac{i}{e^{i\frac{\pi}{2}}|\hbar|}s_{\frac{\pi}{2}}(a_{\alpha})\right)\right].
\end{equation}

\section{TBA equations for WKB periods}
\label{sec:MTBA}

\subsection{TBA equations}
\label{subsec:TBAeq}
To simplify the discontinuities (\ref{eq:adisc1})$\sim$(\ref{eq:bdisc}), we introduce the functions $\epsilon_i(\theta)$ by
\begin{equation}
\label{eq:pseudo}
 \epsilon_{1}\left(\theta + \frac{i\pi}{2} - i\varphi\right) = \frac{i}{\hbar}s_{\varphi}\left(a_{\alpha}\right)\left(\hbar\right),\ \ \ \ \  \epsilon_{2}\left(\theta - i\varphi \right) = \frac{i}{\hbar}s_{\varphi}\left(a_{\beta}\right)\left(\hbar\right),
\end{equation}
where we define $\theta$ by
\begin{equation}
 \frac{1}{\hbar} = e^{\theta - i\varphi}.
\end{equation}
Now we can express the discontinuity formulae in the form:
\begin{equation}
\label{eq:disc}
 \mathrm{disc}_{\pm\pi/2}\epsilon_i\left(\theta\right) = \pm\left[L_{i-1}(\theta) + L_{i+1}(\theta)\right], \ \ \ \ (i = 1, 2)
\end{equation}
where
\begin{equation}
 L_{i}(\theta) = \log\left(1 + e^{-\epsilon_i(\theta)}\right)
\end{equation}
and we define $L_0 = L_2$, $L_3 = L_1$. From the large $\theta$ expansion of (\ref{eq:pseudo}), $\epsilon_i(\theta)$ has the asymptotic behavior 
\begin{equation}
  \epsilon_i(\theta) = m_ie^{\theta} + \mathcal{O}(e^{-\theta}), \ \ \ \theta \rightarrow \infty.
\end{equation}
Now we can find that the functions $\epsilon_i(\theta)$ is given by the following TBA equations \cite{Ito}
\begin{equation}
\label{eq:TBAM2}
\begin{split}
 \epsilon_1(\theta) &= m_1e^{\theta} - 2\int_{\mathbb{R}}\frac{\log\left(1 + e^{-\epsilon_2(\theta')}\right)}{\cosh (\theta - \theta')}\frac{d\theta'}{2\pi}, \\
 \epsilon_2(\theta) &= m_2e^{\theta} - 2\int_{\mathbb{R}}\frac{\log\left(1 + e^{-\epsilon_1(\theta')}\right)}{\cosh (\theta - \theta')}\frac{d\theta'}{2\pi}.
\end{split}
\end{equation}
\par
This TBA equations coincide with the "conformal limit" of the low-energy effective theory of the $\mathcal{N} = 2$ gauge theory on $\mathbb{R}^3 \times S^1$ \cite{GMN,Opers}. We derived the TBA equations by using the discontinuity formula. On the other hand, the TBA equations for the Mathieu equation are also considered in \cite{Fioravanti}, where the authors derive the TBA equations for the periods $a_{\beta}(\hbar, u)$ and $a_{\beta}(\hbar, -u)$ by using the functional relation of the wronskian for the subdominant asymptotic solutions. The derivation in \cite{Fioravanti} is based on the ODE/IM correspondence \cite{TD,BLZ}.
\par
The TBA equations are simplified when we choose $u = 0$. Then we have 
\begin{equation}
 m_1 = m_2 =: m.
\end{equation} 
In this case, from the calculation of (\ref{eq:TBAM2}), we can see that $\epsilon_1$ and $\epsilon_2$ are identified 
\begin{equation}
 \epsilon_1 = \epsilon_2 =: \epsilon.
\end{equation} 
Therefore the two TBA equations collapse to one,
\begin{equation}
\label{eq:TBAMu0}
 \epsilon(\theta) = me^{\theta} - 2\int_{\mathbb{R}}\frac{\log\left(1 + e^{-\epsilon(\theta')}\right)}{\cosh (\theta - \theta')}\frac{d\theta'}{2\pi}.
\end{equation}
This TBA equation coincides with the massless Sinh-Gordon TBA equation, which is the result of the conformal limit of the TBA equation for the Sinh-Gordon model \cite{ShG}. The TBA equation (\ref{eq:TBAMu0}) is also studied in the context of the ODE/IM correspondence for the generalized Mathieu equation \cite{Zad}. 
\par
So far we investigate the TBA equations for real $u$ satisfying $-\Lambda^2 < u < \Lambda^2$. But we can extend the TBA equations for complex $u$ as far as $u$ belongs to the minimal chamber \cite{Grassi,GMN}. On the complex $u$-plane, the minimal chamber coincide with the inside of the marginal stability curve \cite{MS}. The marginal stability curve is real co-dimension one curve in $u$-plane which runs through the points $u = \pm\Lambda^2$. In the inside of the marginal stability curve, the TBA equations are written as follows :
\begin{equation}
 m_1 = |m_1|e^{i\phi_1} = a_{\alpha}^{(0)} , \ \ m_2 = |m_2|e^{i\phi_2} = ia_{\beta}^{(0)}
\end{equation}
\begin{equation}
\label{eq:cTBA}
\begin{split}
 \tilde{\epsilon_1}(\theta) &= |m_1|e^{\theta} - 2\int_{\mathbb{R}}\frac{\log\left(1 + e^{-\tilde{\epsilon_2}(\theta')}\right)}{\cosh (\theta - \theta' + i\phi_2 - i\phi_1)}\frac{d\theta'}{2\pi}, \\
 \tilde{\epsilon_2}(\theta) &= |m_2|e^{\theta} - 2\int_{\mathbb{R}}\frac{\log\left(1 + e^{-\tilde{\epsilon_1}(\theta')}\right)}{\cosh (\theta - \theta' + i\phi_1 - i\phi_2)}\frac{d\theta'}{2\pi},
\end{split}
\end{equation}
where 
\begin{equation}
 \tilde{\epsilon_i}(\theta) := \epsilon_i(\theta -i\phi_i)\ \ \ (i = 1, 2),
\end{equation}
and $|\phi_1 - \phi_2| < \frac{\pi}{2}$. 
\par
The integral equations for the outside of the marginal stability region can be also obtained, but this equation is no longer of the form of the TBA equations and it is difficult to solve even numerically.

\subsection{All-order expansion of the WKB periods}
\label{subsec:allorder}
In this subsection, we study the $\hbar = e^{-\theta}$ expansion of the solution to the TBA equations and compare their with the WKB expansion of the periods. When we expand (\ref{eq:TBAM2}) at large $\theta$, we obtain all-order asymptotic expansion,
\begin{equation}
\label{eq:asym}
  \epsilon_i(\theta) = m_ie^{\theta} + \sum_{n=1}^{\infty}m_i^{(n)}e^{(1-2n)\theta}, \ \ \ \theta \rightarrow \infty,
\end{equation}
where
\begin{equation}
\label{eq:allorder}
 m_k^{(n)} = \frac{(-1)^n}{\pi}\int_{\mathbb{R}} e^{(2n-1)\theta}(L_{k-1}(\theta) + L_{k+1}(\theta))d\theta  \ \ (k = 1, 2).
\end{equation}

\begin{table}[htbp]
\begin{center}
\hspace*{-0.3in} 
\begin{tabular}{ccccc} \hline
n & $(-1)^na_{\alpha}^{(n)}$ & $m_1^{(n)}$ & $ia_{\beta}^{(n)}$ & $m_2^{(n)}$ \\ \hline
0 &                             &  \ 1.82531050  &                       & \ 8.22561885    \\
1 & -\underline{0.15993}792 & -\underline{0.15993}849 & -\underline{0.42667}081 & -\underline{0.42667}143 \\ 
2 & \ \underline{0.00420150} & \ \underline{0.00420150} & \ \underline{0.2239899}8 & \ \underline{0.2239899}9 \\ \hline
\end{tabular}
\caption{The numerical results of the coefficients of the standard WKB periods ($\Lambda = 1$, $u = -0.6$). The numerical calculation in the TBA equations is done by Fourier discretization with $2^{18}$ points and a cutoff of the integrals $(-L, L)$ where $L = 41.837877$.}
\label{tab:compare}
\end{center}
\end{table}

\begin{figure}[htbp]
  \begin{center}
    \begin{tabular}{c}

      \begin{minipage}{0.47\hsize}
        \begin{center}
          \includegraphics[clip, width=6.5cm]{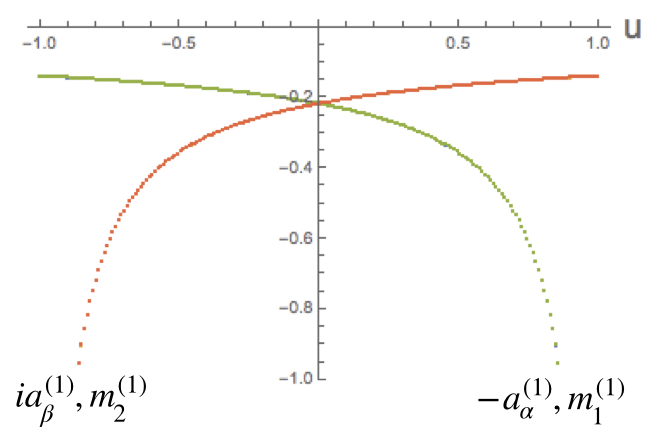}
           \hspace{1.6cm}[a] first order
        \end{center}
      \end{minipage}

      \begin{minipage}{0.47\hsize}
        \begin{center}
          \includegraphics[clip, width=6.5cm]{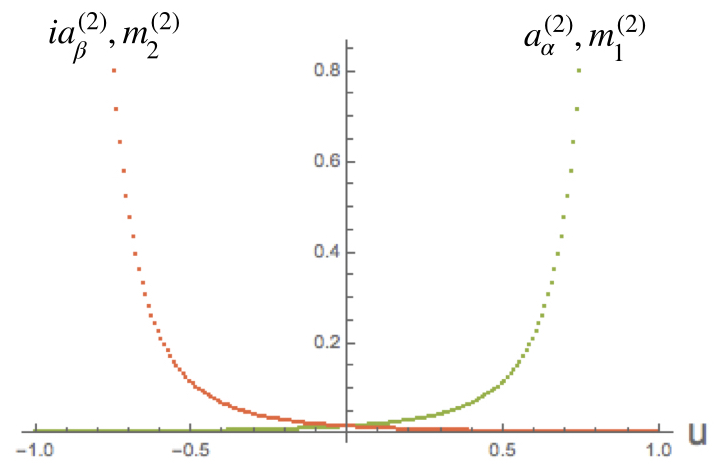}
          \hspace{1.6cm} [b] second order
        \end{center}
      \end{minipage}

    \end{tabular}
    \caption{The numerical results of [a] the first order and [b] the second order ($\Lambda = 1$, $u = -0.6$). The green lines show the results of $(-1)^na_{\alpha}^{(n)}$, $m_1^{(n)}$ and the red lines show the results of $ia_{\beta}^{(n)}$, $m_2^{(n)}$.}
    \label{fig:12coef}
  \end{center}
\end{figure}

From $m_k^{(n)}$, we can recover the coefficients of the standard WKB periods as
\begin{equation}
\label{eq:Eper}
 m_{1}^{(n)} = (-1)^na_{\alpha}^{(n)}, \ \ \ \ \ m_{2}^{(n)} = ia_{\beta}^{(n)}.
\end{equation}
In Table \ref{tab:compare} we compare $m_k^{(n)}$ to $a_{\alpha}^{(n)}$ and $a_{\beta}^{(n)}$ for the first two terms at $u = -0.6$, $\Lambda = 1$. We note that we can choose $\Lambda = 1$ without loss of generality from the quasi-homogeneous property of the Mathieu equation. We also compare the numerical results with $-1 < u < 1$ in Figure \ref{fig:12coef}. To compute the higher order corrections of $a_{\alpha}^{(n)}$ and $a_{\beta}^{(n)}$, we used the differential operator technique\cite{Kashani,He}. The classical periods $a_{\alpha}^{(0)}$ and $a_{\beta}^{(0)}$ for $-1 < |u| < 1$ can be expressed in terms of the hypergeometric function,
\begin{equation}
\label{eq:cl2F1}
\begin{split}
 a_{\alpha}^{(0)} &= \frac{16}{\sqrt{2}}\frac{\pi}{2}\lr{ _2F_1\left(\frac{1}{2}, -\frac{1}{2}, 1, \frac{1+u}{2}\right) - \frac{1}{2}\lr{1-u} _2F_1\left(\frac{1}{2}, \frac{1}{2}, 1, \frac{1+u}{2}\right)}, \\
 a^{(0)}_{\beta} &= -\frac{16i}{\sqrt{2}}\frac{\pi}{2}\lr{ _2F_1\left(\frac{1}{2}, -\frac{1}{2}, 1, \frac{1-u}{2}\right) - \frac{1}{2}\lr{1+u} _2F_1\left(\frac{1}{2}, \frac{1}{2}, 1, \frac{1-u}{2}\right)}.
\end{split}
\end{equation}
The higher order corrections can be computed by the differential operators $D_n$ that act on the classical periods,
\beq
 a_{\gamma}^{(n)} = D_na_{\gamma}^{(0)}.
\eeq
For $\Lambda = 1$, the first few orders of the differential operators can be expressed as follows:
\begin{equation}
\label{eq:Diff}
\begin{split}
 &D_1 = \frac{1}{24}(2u\partial^2_u + \partial_u), \\
 &D_2 = \frac{1}{2^7}(\frac{28}{45}u^2\partial^4_u + \frac{8}{3}u\partial^3_u + \frac{5}{3}\partial^2_u).
\end{split}
\end{equation}

\begin{figure}[htbp]
  \begin{center}
    \begin{tabular}{c}

      \begin{minipage}{0.47\hsize}
        \begin{center}
          \includegraphics[clip, width=5.5cm]{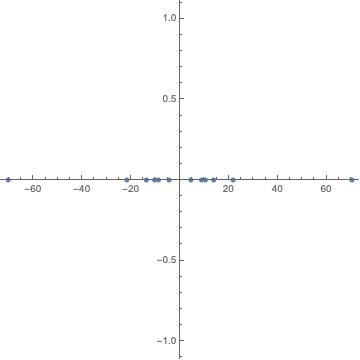}
           \hspace{1.6cm} [a] $a_{\alpha}$
        \end{center}
      \end{minipage}

      \begin{minipage}{0.47\hsize}
        \begin{center}
          \includegraphics[clip, width=5.5cm]{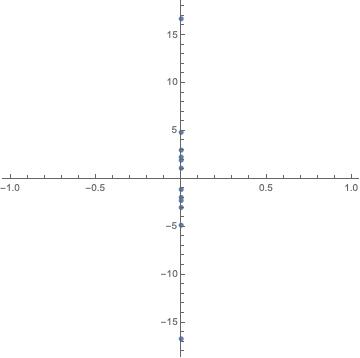}
          \hspace{1.6cm} [b] $a_{\beta}$
        \end{center}
      \end{minipage}

    \end{tabular}
    \caption{The poles of the Borel transformation of [a] $a_{\alpha}$ and [b] $a_{\beta}$ on the complex $\xi$-plane ($\Lambda = 1$, $u = -0.6$).}
    \label{fig:pade}
  \end{center}
\end{figure}

From the TBA equations, we can also calculate higher order collections to the standard periods numerically. In Figure \ref{fig:pade}, we show that the poles of the Borel transformation of the standard periods. To compute the Borel transformations, we use the standard periods which are determined by  (\ref{eq:allorder}) and the diagonal Pad\'{e} approximant of order $12$. These positions of the poles are consistent with the discontinuities (\ref{eq:adisc1})$\sim$(\ref{eq:bdisc}). The poles of the Borel-Pad\'{e} transformations are also calculated in \cite{Grassi}. They use 193 terms of the standard periods which is determined by the differential operator technique. To reproduce their results, we need higher order corrections and more precision of the calculation of (\ref{eq:allorder}).

\subsection{Effective central charge and PNP relation}
\label{subsec:PNP}

In the TBA systems (\ref{eq:TBAM2}), the functions $\epsilon_i(\theta)$ are regarded as the energy of pseudo particles of the integrable field theories. In the UV limit, we can evaluate the "effective central charge" $c_{\mathrm{eff}} = c - 24\Delta_{min}$ of the underlying CFT , where $c$ is the central charge of the Virasoro algebra and $\Delta_{min}$ is the minimum eigenvalue of the Virasoro operator $L_0$. $c_{\mathrm{eff}}$ is given by
\begin{equation}
\label{eq:ceff}
  c_{\mathrm{eff}} = \frac{6}{\pi^2}\sum_{i = 1}^{2}m_i\int_{\mathbb{R}}e^{\theta}L_i(\theta)d\theta = 2 + \frac{3}{\pi^2}\sum_{i=1}^{2}\left(\epsilon_i^{\star}\log(1+e^{\epsilon_i^{\star}}) + 2{\mathrm{Li}}_2(-e^{\epsilon_i^{\star}})\right),
\end{equation}
where 
\beq
\epsilon_i^{\star} = \lim_{\theta \rightarrow -\infty}\epsilon_i(\theta).
\eeq
We calculate the effective central charge for the TBA equations (\ref{eq:TBAM2}). The $\theta \rightarrow -\infty$ limit of the pseudo-energies for (\ref{eq:TBAM2}) satisfy
\begin{equation}
 e^{-\epsilon_1^{\star}} = 1 + e^{-\epsilon_2^{\star}} , \ \ \ e^{-\epsilon_2^{\star}} = 1 + e^{-\epsilon_1^{\star}}.
\end{equation} 
In \cite{Opers,Cec}, it is pointed out that there are no mathematically rigorous solutions to these equations. But we can formally consider that $\epsilon_i^{\star} \rightarrow -\infty$ are the solutions to these equations. Then we can identify the effective central charge associated with TBA equations (\ref{eq:TBAM2}),
\begin{equation}
  c_{\mathrm{eff}} = 2.
\end{equation}
This result agrees with the numerical calculation.
\par
We can also compute the effective central charge of the second term of (\ref{eq:ceff}) from the standard WKB periods. The large $\theta$ expansion for (\ref{eq:cTBA}) leads to
\begin{equation}
\label{eq:periodconst}
  c_{\mathrm{eff}} = -\frac{3i}{\pi}\left(a_{\alpha}^{(0)}a_{\beta}^{(1)} - a_{\alpha}^{(1)}a_{\beta}^{(0)}\right).
\end{equation}
We can evaluate this by the hypergeometric form of the standard WKB periods, which is again equal to 2. Note that the classical periods (\ref{eq:cl2F1}) satisfy the second-order Picard-Fuchs equation \cite{Ceresole},
\begin{equation}
\label{eq:PF}
 \frac{\partial^2}{\partial u^2}\left(a_{\alpha}^{(0)}, a_{\beta}^{(0)}\right) = \frac{-1}{4(u+1)(u-1)}\left(a_{\alpha}^{(0)}, a_{\beta}^{(0)}\right).
\end{equation}
This equation allows us to rewrite (\ref{eq:periodconst}) as
\begin{equation}
\label{eq:diff}
 -\frac{3i}{\pi}\left(a_{\alpha}^{(0)}a_{\beta}^{(1)} - a_{\beta}^{(1)}a_{\alpha}^{(0)}\right) = -\frac{i}{8\pi}\left(a_{\alpha}^{(0)}\frac{\partial a_{\beta}^{(0)}}{\partial u} - \frac{\partial a_{\alpha}^{(0)}}{\partial u}a_{\beta}^{(0)}\right).
\end{equation}
From the wronskian relation \cite{Matone} satisfied by the classical periods
\begin{equation}
 a_{\alpha}^{(0)}\frac{\partial a_{\beta}^{(0)}}{\partial u} - \frac{\partial a_{\alpha}^{(0)}}{\partial u}a_{\beta}^{(0)} = 16\pi i,
\end{equation}
we obtain
\begin{equation}
 -\frac{3i}{\pi}\left(a_{\alpha}^{(0)}a_{\beta}^{(1)} - a_{\beta}^{(1)}a_{\alpha}^{(0)}\right) = 2.
\end{equation}
This non-trivial relation between the independent standard periods is called the PNP-relation. The PNP-relation also exists for several classes of the potentials \cite{HPNP}, \cite{ChePNP}.
\vspace{0.1in}
\par
In \cite{MM}, it is claimed that the standard WKB periods $a_{\alpha}$, $a_{\beta}$ are equivalent to the quantum periods for 4-dimensional $\mathcal{N} = 2$ $SU(2)$ super Yang-Mills theory and explicitly demonstrated up to the lowest orders. Therefore the TBA equations (\ref{eq:cTBA}) imply that the Borel resummation of the quantum periods, which are the quantities of the 4-dimensional QFT, are equal to the pseudo-energies, which is the quantities of the 2-dimensional CFT, at least on the inside of the marginal stability. Moreover, in \cite{EY,STY}, it is show that the r.h.s of (\ref{eq:diff}) is proportional to the coefficient of the one-loop beta function for the $\mathcal{N} = 2$ theory. Therefore (\ref{eq:periodconst}) indicates that the quantum collection to the beta function for the $\mathcal{N} = 2$ theory is governed by the effective central charge of the 2-dimensional CFT.

\section{Spectral problem for the Mathieu equation}
\label{sec:band}
The energy spectrum of the Mathieu equation has the band structure. In \cite{ZJJ}, Zinn-Justin and Jentschura have conjectured that the exact quantization condition for the periodic cosine potential is given by the following equation, 
\beq
\label{eq:ZJJ}
 \lr{\frac{32}{\sqrt{2}\hbar}}^{-B_{\mathrm{ZJJ}}}\frac{e^{\frac{A_{\mathrm{ZJJ}}}{2}}}{\Gamma\lr{\frac{1}{2} - B_{\mathrm{ZJJ}}}} + \lr{-\frac{32}{\sqrt{2}\hbar}}^{B_{\mathrm{ZJJ}}}\frac{e^{\frac{-A_{\mathrm{ZJJ}}}{2}}}{\Gamma\lr{\frac{1}{2} + B_{\mathrm{ZJJ}}}} = \frac{2\cos\theta_{\mathrm{B}}}{\sqrt{2\pi}},
\eeq

\begin{figure}[htbp]
  \begin{center}
   \includegraphics[width=160mm]{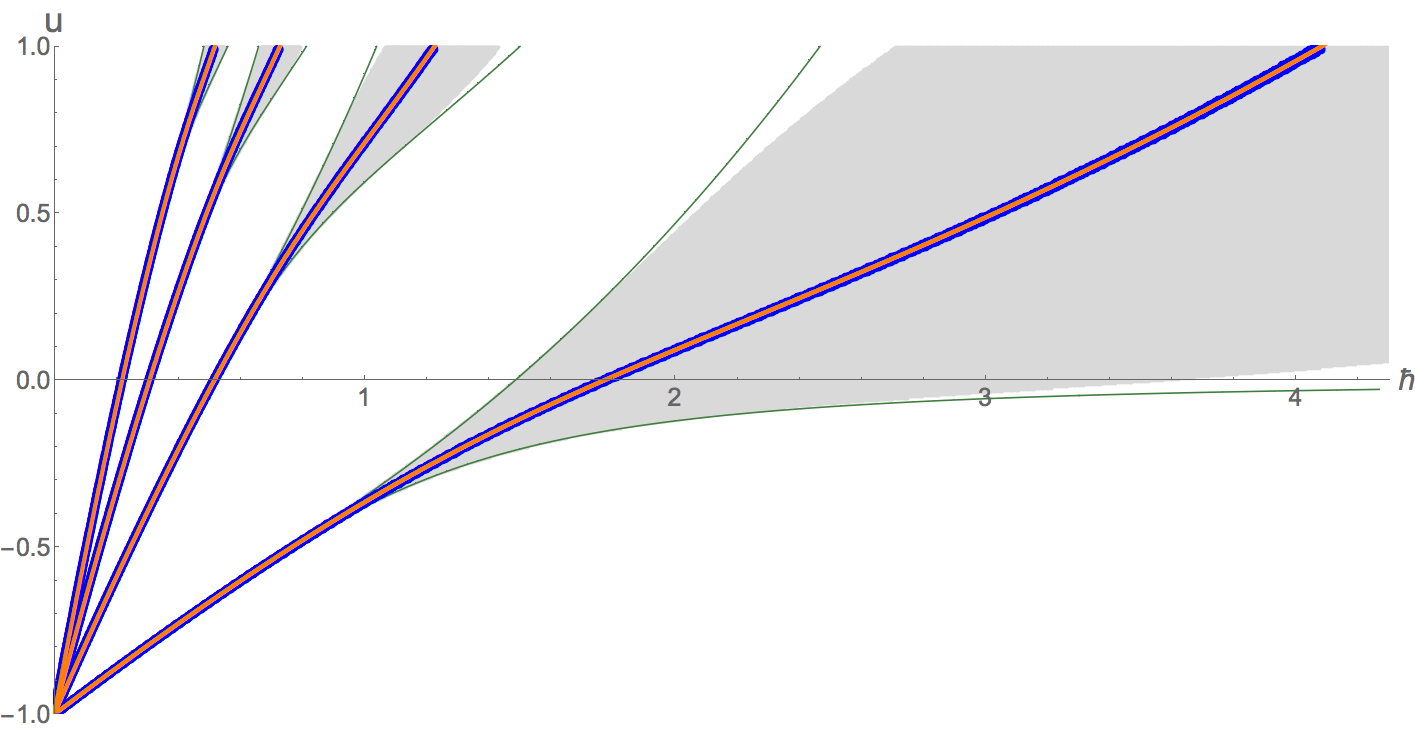}
   \caption{(i) The exact energy bands (green lines). (ii) The energy spectrum determined by the Bohr-Sommerfeld condition for $s_{med}\lr{a_{\alpha}}\lr{u, \hbar}$ (blue lines). (iii) The exact spectrum at $\theta_{\mathrm{B}} = \frac{\pi}{2}$ (orange lines). (iv) The energy bands determined by (\ref{eq:ZJJ}) with the exact periods (gray region). The numerical calculation in the TBA equations is done by taking the value $\delta = 10^{-15}$.}
   \label{fig:BP}
  \end{center}
\end{figure}

\hspace*{-0.4in} where $\theta_{\mathrm{B}}$ is the Bloch angle. $A_{\mathrm{ZJJ}}$ and $B_{\mathrm{ZJJ}}$ are formal power series in $\hbar$ and related to the standard WKB periods as follows \cite{BD}:
\beq
 B_{\mathrm{ZJJ}}\lr{u, \hbar} = \frac{1}{2\pi \hbar}a_{\alpha}\lr{u, \hbar},
\eeq
\beq
 A_{\mathrm{ZJJ}}\lr{u, \hbar} = \frac{i}{\hbar}a_{\beta}\lr{u, \hbar} -2\ln\Gamma\lr{\frac{1}{2} + B_{\mathrm{ZJJ}}\lr{u, \hbar}} + \ln\lr{2\pi} - 2B_{\mathrm{ZJJ}}\lr{u, \hbar}\ln\lr{\frac{\sqrt{2}\hbar}{32}}.
\eeq
In particular, the perturbative contribution to the energy spectrum can be obtained by the following condition,
\beq
\label{eq:BS}
 B_{\mathrm{ZJJ}}\lr{u, \hbar} = N + \frac{1}{2}, \ \ \ (N \in \mathbb{N}_{\geq 0}).
\eeq
This equation corresponds to the Bohr-Sommerfeld quantization condition for $a_{\alpha}$.
\par
The quantization conditions (\ref{eq:ZJJ})$\sim$(\ref{eq:BS}) are only valid for $u \ll 1$ and $\hbar \ll 1$, in which we can use the approximation by the standard WKB periods. Instead of the standard periods, we will use the exact WKB periods $s_{med}\lr{a_{\alpha}}$, $s_{0}\lr{a_{\beta}}$, where $s_{med}$ indicates the median resummation. The median resummation can be easily computed as the principal value of the singular integral \cite{Ito},
\beq
 \frac{1}{\hbar}s_{med}\lr{a_{\alpha}}\lr{u, \hbar} = m_1e^{\theta} + \mathrm{P}\int_{\mathbb{R}}\frac{2 L_2(\theta)}{\sinh{(\theta - \theta')}}\frac{d\theta'}{2\pi},
\eeq
where
\beq
 \mathrm{P}\int_{\mathbb{R}}\frac{2 L_2(\theta)}{\sinh{(\theta - \theta')}}\frac{d\theta'}{2\pi} := \lim_{\delta \rightarrow 0} \int_{\mathbb{R}}\frac{2L_2(\theta')\sinh{(\theta - \theta')\cos{\delta}}}{\sinh^2{(\theta - \theta')}\cos^2{(\delta)} + \cosh^2{(\theta - \theta')}\sin^2{(\delta)}}\frac{d\theta'}{2\pi}.
\eeq

In Fig.\ref{fig:BP}, we depict the energy spectrum with several computations. The gray region indicates the energy bands calculated by (\ref{eq:ZJJ}) with the exact periods. We can see that the gray region partially reproduce the exact band spectrum (green lines). Thus the exact WKB periods extend the valid region of the quantization condition (\ref{eq:ZJJ}). The blue lines show the results of the Bohr-Sommerfeld condition for $s_{med}\lr{a_{\alpha}}\lr{u, \hbar}$. And the orange lines are the exact spectrum at $\theta_{\mathrm{B}} = \frac{\pi}{2}$, which includes the two and much higher instanton contributions but approximately determines the exact perturbative spectrum. 

\section{Conclusions and discussions}
\label{sec:CandD}
In this paper, we derived the TBA equations which govern the exact WKB periods for the Mathieu equation at weak coupling. The TBA equations gave an efficient way to calculate not only the exact periods but also the all-order coefficients for the standard periods. The TBA equations also provided some new perspectives for the related subjects. First, we have seen that the coefficients of the one-loop beta function for 4-dimensional $\mathcal{N} = 2$ $SU(2)$ super Yang-Mills theory is proportional to the effective central charge for the 2d CFT. Second, in section \ref{sec:band}, the exact WKB period with the Bohr-Sommerfeld quantization condition determined the perturbative contribution to the energy spectrum at least $-\Lambda^2 < u < \Lambda^2$. 
\par
Our results raise several open problems. In section \ref{sec:band}, the bandwidth determined by the TBA equations were not agreed with the exact results as $u$ or $\hbar$ become large. To compute the energy spectrum for these regions, we may be necessary to consider the exact quantization conditions for the Mathieu equation by using the exact WKB method \cite{Esemi}. It is also interesting to apply our method to other periodic potentials and study their band structure. The Hill's differential equation is one of them, which includes the Mathieu equation as a special case.
\par
There are also many crucial generalizations in the context of gauge theories. First of all, whether we can apply our method to the theories with matters or not. These theories have more complicated forms of the potentials (see e.g.\cite{IOK}). Another direction for generalization is $\mathcal{N} = 2^{\ast}$ $SU(2)$ super Yang-mills theory, whose quantum periods can be calculated from the Lam\'{e} equation \cite{BD,N2star,Lam}.
\par
The most important problem is the formulation for the outside of the strong coupling region. As stated in subsection \ref{subsec:TBAeq}, integral equations in the outside of marginal stability have no longer the form of TBA equations. On the other hand, in \cite{T3}, other type of integral equations for these regions are derived by using the abelianization technique. They use a form of the Mathieu equation which can be obtained from the change of variable $e^{iq} = z$. We probably need the exact WKB analysis in this coordinate \cite{Troost}.

\section*{Acknowledgements}
We would like to thank Katsushi Ito, Kohei Kuroda, Takayasu Kondo, Saki Koizumi, Hongfei Shu, Yasuyuki Hatsuda, Reona Arai for very useful discussions. We are particularly grateful to Katsushi Ito and Yasuyuki Hatsuda for the useful advice in the preparation of this paper. We also acknowledge the financial support from Advanced Research Center for Quantum Physics and Nanoscience, Tokyo Institute of Technology.

\end{document}